\documentclass[a4paper,11pt]{article}
\pdfoutput=1 

\usepackage{jheppub} 

\usepackage[T1]{fontenc} 
\usepackage{slashed}
\usepackage{multirow}

\title{Quarkonium light-cone distribution amplitudes: twist structure and mass dependence}

\author[a]{Shuai Xu,}
\author[b,1]{Xiao-Nan Li,\note{Corresponding author.}}
\author[a]{Jin-Zhong Han,}
\author[a]{Bai-Hui Cheng,}
\author[c]{Li-Li Chen,}
\author[c]{Qin Chang}

\affiliation[a]{School of Physics and Telecommunications Engineering, Zhoukou Normal University, \\Zhoukou 466001, P.R. China}
\affiliation[b]{School of Electrical and Information Engineering, Tongling University, \\Tongling 244061, P.R. China}
\affiliation[c]{Institute of Particle and Nuclear Physics, Henan Normal University, \\Xinxiang 453007, P.R. China}

\emailAdd{xushuai@zknu.edu.cn}
\emailAdd{lixn@tlu.edu.cn}
\emailAdd{hanjinzhong@zknu.edu.cn}
\emailAdd{cbh@zknu.edu.cn}
\emailAdd{chenlili2020@htu.edu.cn}
\emailAdd{changqin@htu.edu.cn}

\abstract{
We present a systematic study of the leading- and next-to-leading-twist light-cone distribution amplitudes (LCDAs) of ground-state pseudoscalar and vector quarkonium within the light-front quark model (LFQM). By implementing the replacement $M \to M_0$, we analyze the longitudinal and transverse structures of the LCDAs, together with their Gegenbauer moments, $\xi$-moments, and transverse momentum moments. We show that charge-conjugation symmetry enforces the exact vanishing of all odd Gegenbauer moments and odd $\xi$-moments. For pseudoscalar quarkonium, the twist-2 and twist-3 LCDAs become identical, which leads to the same Gegenbauer moments, $\xi$-moments, and transverse momentum moments. For vector quarkonium, although the twist-2 and twist-3 LCDAs differ in the case of finite quark masses, they progressively converge as the quark mass increases. In the heavy-quark limit, all quarkonium LCDAs satisfy $\phi^A_{2} = \phi^P_{3} \simeq \phi^{\parallel}_{2} \simeq \phi^{\perp}_{3}$,
demonstrating an emergent twist-independence of quarkonium distribution amplitudes. We further find that the LCDAs become increasingly peaked and narrower with increasing quark mass, indicating that the meson system becomes increasingly close to a nonrelativistic bound state, with a more uniform and stable distribution of internal longitudinal momentum. For all quarkonium, the peak value exhibits a simple phenomenological scaling behavior governed by the ratio $m/\beta$. The transverse momentum moments increase with the meson mass, indicating a progressively more compact bound-state structure. These results reveal a universal and systematic evolution of quarkonium LCDAs driven by the quark mass.
}

\begin{document}
\maketitle
\flushbottom

\section{Introduction}
\label{sec:intro}

LCDAs encode essential nonperturbative information of hadrons and play a central role in the QCD description of hard exclusive processes. As universal objects, LCDAs describe the longitudinal momentum distributions of valence constituents at fixed light-cone separation and serve as indispensable inputs for QCD factorization approaches. Understanding their structural properties, twist-dependence, and mass evolution is therefore crucial for connecting nonperturbative dynamics with high-energy observables.

Quarkonium systems, composed of a quark and its antiquark, provide a particularly clean and theoretically controlled system for investigating the internal structure of hadrons. Owing to charge-conjugation symmetry and the presence of a well-defined heavy-quark limit, quarkonium exhibits an intrinsically nonrelativistic character, in which spin-dependent and relativistic effects are suppressed~\cite{Quigg:1979vr}. These features make quarkonium an ideal laboratory for exploring how the twist structure of LCDAs evolves with the quark mass and for addressing a fundamental question: to what extent the distinction between different twists remains physically relevant in the heavy-quark limit.

Although non-relativistic QCD (NRQCD)~\cite{Bodwin:1994jh,Bodwin:2006dn}
is a powerful theoretical tool for separating high-energy modes from low-energy contributions, in most cases the calculation of low-energy hadronic matrix elements has relied
on model-dependent nonperturbative methods. In recent years, substantial progress has been made in the determination of LCDAs using various nonperturbative approaches, including QCD sum rules~\cite{Chernyak:1983ej,Colangelo:2000dp,Bakulev:2005cp,Ball:1998sk,Yang:2007zt,Zhang:2025qmg,Zeng:2025rfe,Wang:2025sic,Han:2013zg,Zhong:2021epq,Khodjamirian:2004ga,Khodjamirian:2020hob,Braguta:2008qe,Braguta:2007tq,Braguta:2006wr,Braguta:2007fh}, lattice QCD~\cite{CP-PACS:2001vqx,Braun:2006dg,LatticeParton:2024vck,LatticeParton:2024zko,LatticeParton:2022zqc,Ding:2024saz,Baker:2024zcd,Blossier:2024wyx,Zhang:2020gaj,Cloet:2024vbv}, perturbative QCD (pQCD)~\cite{Lepage:1980fj,Chai:2025xuz,Cheng:2019ruz,Cheng:2020vwr}, Dyson--Schwinger equations~\cite{Maris:1997tm,Roberts:1994dr,Chang:2013pq,Chang:2013epa,Shi:2015esa,Roberts:2021nhw,Xu:2025hjf,Ding:2015rkn}, chiral quark models~\cite{Petrov:1998kg,Nam:2006au,Son:2024uet,Broniowski:2007si}, Nambu-Jona-Lasinio models~\cite{RuizArriola:2002bp,Praszalowicz:2001wy,Noguera:2015iia,Courtoy:2019cxq}, and the  LFQM~\cite{Choi:2007yu,Hwang:2008qi,Ji:1992yf,Brodsky:1997de,Terentev:1976jk,Chen:2021ywv,Chang:2020wvs}. While these studies have provided valuable insights into individual mesons and specific twists, a systematic and unified understanding of the longitudinal and transverse structures of quarkonium LCDAs---covering both pseudoscalar and vector states across different twists---remains incomplete, especially regarding their behavior and possible universality in the heavy-quark limit.

The LFQM offers a powerful framework for addressing this issue\cite{Jaus1, CCH1, Jaus2, CCH2, Hwang, Wei,Hwang:2009cu,Chang:2019mmh,Choi:2017uos,Dhiman:2019ddr,Arifi:2025olq,Choi:2013mda}. By formulating hadron structure in terms of light-front wave functions with explicit transverse momentum dependence, the LFQM provides direct access to both longitudinal and transverse dynamics in a manifestly relativistic framework. Moreover, the replacement of the physical meson mass $M$ by the invariant mass $M_0$ of the constituent quark-antiquark system has been shown to restore self-consistency and covariance in the LFQM, and has led to successful descriptions of nonperturbative quantities, such as decay constants, form factors, and charge radii of mesons\cite{Jaus1, CCH1, Hwang,Chang:2019obq,Choi:2025bxk,Zhang:2025pde,Xu:2025aow,Xu:2025ntz,Li:2026wmb,Lih:2025cmf,Lih:2026xgi,Wang:2024mjw,Hsiao:2020gtc,Arifi:2022pal}. The self-consistent LFQM is based on the Bakamjian--Thomas (BT) construction~\cite{Bakamjian:1953kh,Chung:1988my,Polyzou:2010zz}, in which a meson state is described by an on-shell quark--antiquark pair and interactions are incorporated through the mass operator $M = M_0 + V$, where $M_0$ is the internal invariant mass constructed from the light-front variables $(x, k_\perp)$. It is defined as $M_0^2 = (m^2_q + k_\perp^2)/x+ (m^2_{\bar{q}} + k_\perp^2)/\bar{x}$. The invariant mass $M_0$ plays a crucial role in the analysis of covariance and self-consistency between the standard and covariant formulations of LFQM~\cite{Chang:2019obq,Chang:2019mmh,Choi:2017uos}. Under the variable transformation $\{x, k_\perp\} \to \tilde{k} \equiv \{k_\perp, k_z\}$, the longitudinal momentum is defined as $k_z = \left(x - 1/2\right) M_0 $, with the corresponding Jacobian $\frac{\partial k_z}{\partial x}=\frac{M_0}{4 x \bar{x}}$. At the meson-quark vertex, light-front energy conservation requires $P^- = p_q^- + p_{\bar q}^- $. This condition implies that the external meson mass entering the integrand requires consistency with invariant mass $M_0(x, k_\perp)$. Otherwise, the kinematical relation $(M^2 + P_\perp^2)/P^+=(m_q^2 + p_{q\perp}^2)/p_q^++(m_{\bar q}^2 + p_{\bar q\perp}^2)/{p_{\bar q}^+}$ would be violated. Therefore, the replacement $M \to M_0$ follows from enforcing light-front energy conservation within the BT framework, and is required for maintaining covariance and self-consistency.

In this work, we perform a systematic investigation of the leading- and next-to-leading-twist
LCDAs of ground-state pseudoscalar and vector quarkonium within the LFQM.
By implementing the replacement $M \to M_0$, we analyze not only the shapes of the distribution
amplitudes but also their Gegenbauer moments, $\xi$-moments, and transverse momentum moments.
Our analysis reveals several noteworthy and conceptually important features. First, for pseudoscalar quarkonium, the replacement $M \to M_0$ leads to an exact identity
of the twist-2 and twist-3 LCDAs, implying that their longitudinal and transverse momentum
structures coincide. Second, for vector quarkonium, although the twist-2 and twist-3 LCDAs are
distinct at finite quark mass, they progressively converge as the constituent quark mass
increases. Remarkably, in the heavy-quark limit, $\phi^A_{2}\simeq \phi^{\parallel}_{2}$, and all quarkonium LCDAs approach the approximate relation
$\phi^A_{2} = \phi^P_{3} \simeq \phi^{\parallel}_2 \simeq \phi^{\perp}_3$, demonstrating an emergent
twist-independence of quarkonium distribution amplitudes. It should be noted that this behavior could be understood as a model-dependent feature within the self-consistent LFQM, rather than a general prediction of QCD. We emphasize that this emergent
twist-independence is not imposed by hand, but arises dynamically from the combined effects of
charge-conjugation symmetry, the replacement $M \to M_0$ and the heavy-quark limit.Furthermore, we find that the longitudinal momentum distribution becomes
increasingly localized around $x=1/2$ as the quark mass grows, indicating a systematic narrowing
of the LCDAs. For pseudoscalar quarkonium, the peak value of the LCDAs exhibits a simple
phenomenological scaling behavior controlled primarily by the ratio $m/\beta$ within our model. The transverse momentum moments show a complementary trend, increasing with
both the moment order and the meson mass, consistent with a progressively more compact spatial
structure.

The remainder of this paper is organized as follows. In Sec.~\ref{sec:2}, we briefly review the LFQM formalism and derive the expressions for quarkonium LCDAs. Sec.~\ref{sec:3} presents our numerical results and a detailed discussion of the distribution amplitudes, Gegenbauer moments, $\xi$-moments, and transverse momentum moments. The summaries are given in Sec.~\ref{sec:4}.

\section{Preliminary} \label{sec:2}
In the LFQM, the Fock state is treated as in a noninteraction $q\bar{q}$ representation and the interaction is encoded in the light-front (LF) wave function $\Psi_{h\bar{h}}(x,\mathbf{k}_\perp)$. Specifically, the expansion for a meson is given by
\begin{equation}
|M(P)\rangle = \int \{d^3p_q\}\{d^3p_{\bar{q}}\}2(2\pi)^3\delta^3(P-p_q-p_{\bar{q}}) \sum_{h,\bar{h}}\Psi_{h\bar{h}}(x,\mathbf{k}_\perp)|q(p_q,h)\bar{q}(p_{\bar{q}},\bar{h})\rangle,
\label{eq:1}
\end{equation}
where $p_q (p_{\bar{q}})$ and $h (\bar{h})$ is momentum and helicity of the constituent quark (antiquark), respectively. The momentum assignments in terms of LF variable $(x,\mathbf{k}_\perp)$ for constituents are as follows
\begin{align}
p^+_q &= x P^+, & p^+_{\bar{q}} &= \bar{x} P^+, \\
\mathbf{p}_{q \perp} &= x\mathbf{P}_\perp+\mathbf{k}_\perp, & \mathbf{p}_{\bar{q} \perp} &= \bar{x}\mathbf{P}_\perp-\mathbf{k}_\perp,
\label{eq:2}
\end{align}
where $p_q^+$ and $\mathbf{p}_{q \perp}$ are longitudinal and transverse momentum. In which, $\bar{x}=1-x$ and $P=p_q+p_{\bar{q}}$ is inherently satisfied. We work in a reference frame where $\mathbf{P}_\perp=0$, such that the transverse momenta of the quark and antiquark reduce to $\mathbf{p}_{q \perp}=\mathbf{k}_\perp$ and $ \mathbf{p}_{\bar{q} \perp}=-\mathbf{k}_\perp$, respectively. The LF wave function $\Psi_{h\bar{h}}(x,\mathbf{k}_\perp)$ is generally defined as
\begin{equation}
\Psi_{h\bar{h}}(x,\mathbf{k}_\perp) = \psi(x,\mathbf{k}_\perp)S_{h\bar{h}}(x,\mathbf{k}_\perp),
\label{eq:3}
\end{equation}
where $\psi(x,\mathbf{k}_\perp)$ is the radial wave function and $S_{h\bar{h}}(x,\mathbf{k}_\perp)$ corresponds to the spin wave function.
For pseudoscalar mesons, a kind of 1S state radial wave function $\psi(x,\mathbf{k}_\perp)$ has been suggested in previous works~\cite{Chang:2019mmh,Choi:2017uos} and the well-proven Gaussian-type wave function is taken in this work
\begin{equation}
\psi(x,{\bf{k}}_{\perp})= \frac{4\pi^{3/4}}{\beta^{3/2}}\sqrt{\frac{\partial k_z}{\partial x}} \exp\!\left(-\frac{{\bf{k}}_{\perp}^2+k_z^2}{2\beta^2}\right),
\label{eq:4}
\end{equation}
where the parameter $\beta$ is meson-dependent and related to the size of the bound state. With the confinements from mesonic decay constants, the values of $\beta$ could be well fixed and the details are presented in the literature~\cite{Xu:2025ntz}. For quarkonium systems in this work, the quark and antiquark masses are equal,
\( m_q = m_{\bar q} = m \). 
For LF spin-orbit wave function, $S_{h\bar{h}}(x,\mathbf{k}_\perp)$ is obtained by the interaction independent Melosh transformation from the traditional spin-orbit wave function and the covariant form is given by
\begin{align}
S_{h\bar{h}}(x,\mathbf{k}_\perp)= \frac{\bar{u}(p_q,h)\Gamma v(p_{\bar{q}},h)}{\sqrt{2}M_0},
\label{eq:7}
\end{align}
where
\begin{align}
\Gamma &= \gamma_5 \quad \text{(for pseudoscalar}\text{)}, \label{eq:Gamma_pseudo}\\
\Gamma &= -\slashed{\epsilon}+ \frac{\epsilon\cdot(p_q-p_{\bar{q}})}{M_0+2m} \quad \text{(for vector}\text{)}, \label{eq:Gamma_vector}
\end{align}
the $u$, $v$ are Dirac spinors and $\sum_{h,\bar{h}}S^\dagger S=1$.

The quark field could be expanded in terms of creation and annihilation operator as
\begin{align}
q(x)&=\int\frac{dp_q^+}{\sqrt{2p_q^+}}\frac{d^2\mathbf{p}_{q\perp}}{(2\pi)^3}\sum_h[b_h(p_q^+,\mathbf{p}_{q\perp})u_h(p_q^+,\mathbf{p}_{q\perp})
e^{-ip_q\cdot x}\nonumber\\
&+d_h^\dagger(p_q^+,\mathbf{p}_{q\perp})v_h(p_q^+,\mathbf{p}_{q\perp})e^{ip_q\cdot x}]
\label{eq:8}
\end{align}
and the expression for $\bar{q}(x)$ can be obtained by taking the conjugate of eq.~(\ref{eq:8}).
With the expressions of meson and quark (antiquark) field in eqs.~(\ref{eq:1},\ref{eq:8}), the matrix elements for $\langle 0|\bar{q}(z)\Gamma q(-z)|M(P)\rangle$ are derived as
\begin{eqnarray}
\mathcal{M}_{\Gamma}=\sqrt{N_c}\sum_{h,\bar{h}}\int\frac{dp_q^+}{\sqrt{2p_q^+}}\frac{d^2\mathbf{p}_\perp}{(2\pi)^3}
\Psi_{h,\bar{h}}(x,\mathbf{k}_\perp)\bar{v}_{\bar{h}}\Gamma u_h e^{i(2x-1)P\cdot z}.
\label{eq:9}
\end{eqnarray}

The DAs of pseudoscalar and vector quarkonium are defined in terms of the following matrix elements
\begin{eqnarray}
\langle 0|\bar{q}(z)\gamma^\mu\gamma^5 q(-z)|M(P)\rangle = if_P P^\mu\int_0^1 dx~e^{i\xi P\cdot z}\phi_{2}^A(x,\mu),\label{eq:10a} \\
\langle 0|\bar{q}(z)i\gamma^5 q(-z)|M(P)\rangle = f_P\mu_M\int_0^1 dx~e^{i\xi P\cdot z}\phi_{3}^P(x,\mu),
\label{eq:10}
\end{eqnarray}
and
\begin{equation}
\langle 0|\bar{q}(0)\gamma^+ q(z^-)|V(P,l)\rangle = f_V M \epsilon_0^+\int_0^1 dx~e^{-i x P\cdot z}\phi_{2}^\parallel(x,\mu),
\label{eq:12}
\end{equation}
\begin{equation}
\langle 0|\bar{q}(0)\gamma^\perp q(z^-)|V(P,l)\rangle = f_V M \epsilon_+^\perp\int_0^1 dx~e^{-i x P\cdot z}\phi_{3}^\perp(x,\mu),
\label{eq:13}
\end{equation}
for $\phi_{3}^\perp(x)$ by taking the perpendicular component $\mu=\perp$ of the current and the transverse polarization $l=+$, respectively.
$P$ denotes the 4-momentum of the meson and the path-ordered gauge link for the gluon fields between the point $-z$ and $z$ is taken. The integration variable $x$ is the longitudinal momentum fraction of the quark and $\xi=x-\bar{x}=2x-1$ depicts longitudinal separation. By taking the Fourier transform for eqs.~(\ref{eq:10a},\ref{eq:12}) with the redefined variable $z^{\mu}=\tau\eta^{\mu}$ where the lightlike vector $\eta=(1,0,0,-1)$, we can obtain
{\small
\begin{eqnarray}
\int_{-\infty}^\infty d\tau \langle 0|\bar{q}(\tau\eta)\gamma^\mu\gamma^5 q(-\tau\eta)|M(P)\rangle e^{-i\xi'\tau P\cdot \eta}=
if_P P^\mu\int_{-\infty}^\infty d\tau \int_0^1 dx~\phi_{2}^A(x,\mu)e^{i(\xi-\xi')\tau P\cdot \eta},
\label{eq:14a}\\
\int_{-\infty}^\infty d\tau \langle 0|\bar{q}(0)\gamma^+ q(-\tau\eta)|V(P,l)\rangle e^{-i\xi'\tau P\cdot \eta}=f_V M \epsilon_0^+\int_{-\infty}^\infty d\tau \int_0^1 dx~\phi_{2}^\parallel(x,\mu)e^{i(\xi-\xi')\tau P\cdot \eta},
\label{eq:14b}
\end{eqnarray}}
in which $\xi'=2x'-1$ with dummy variable $x'$ corresponds to the conjugate variables for $\tau$. Substituting eq.~(\ref{eq:9}) into left hand side of eqs.~(\ref{eq:14a},\ref{eq:14b}), we can directly extract $\phi_{2}^A$($\phi_{2}^\parallel(x,\mu)$) as
\begin{align}
\phi_{2}^A(x,\mu)=\frac{1}{if_P P^\mu}\sqrt{N_c}\sum_{h,\bar{h}}\int\frac{dp_q^+}{\sqrt{2p_q^+}}\frac{d^2\mathbf{p}_\perp}{(2\pi)^3}
\Psi_{h,\bar{h}}(x,\mathbf{k}_\perp)\bar{\nu}_{\bar{h}}\gamma^\mu\gamma^5 u_h,
\label{eq:15a}\\
\phi_{2}^\parallel(x,\mu)=\frac{1}{f_VM \epsilon_0^+}\sqrt{N_c}\sum_{h,\bar{h}}\int\frac{dp_q^+}{\sqrt{2p_q^+}}\frac{d^2\mathbf{p}_\perp}{(2\pi)^3}
\Psi_{h,\bar{h}}(x,\mathbf{k}_\perp)\bar{\nu}_{\bar{h}}\gamma^+ u_h,
\label{eq:15b}
\end{align}
Applying the LF wave function and performing the spinor contraction with $\bar{u}u=\slashed{p}$, we simplify the right hand side of eqs.~(\ref{eq:15a},\ref{eq:15b}). The calculation of the trace term is rather straightforward, from which the twist-2 DAs of the quarkonium follows,

\begin{align}
\phi_2^A(x,\mu) &= \frac{\sqrt{2N_c}}{f_P}\int_0^{\mathbf{k}^2_\perp<\mu^2} \frac{d^2\mathbf{k}_\perp}{8\pi^3}\frac{m}{\sqrt{m^2
+\mathbf{k}^2_\perp}}\psi(x,{\bf{k}}_{\perp}),
\label{eq:16a}\\
\phi_{2}^\parallel(x,\mu) &= \frac{\sqrt{2N_c}}{f_V}\int_0^{\mathbf{k}^2_\perp<\mu^2} \frac{d^2\mathbf{k}^2_\perp}{8\pi^3}\frac{\psi(x,{\bf{k}}_{\perp})}{\sqrt{m^2
+\mathbf{k}^2_\perp}}[m+\frac{2\mathbf{k}^2_\perp}{M_0+2m}].
\label{eq:16b}
\end{align}
Adopting the same procedure, the formula of twist-3 DAs are derived as
\begin{align}
\phi_{3}^P(x,\mu)&=\frac{\sqrt{2N_c}}{f_P\mu_M}\int_0^{\mathbf{k}^2_\perp<\mu^2} \frac{d^2\mathbf{k}_\perp}{16\pi^3}\frac{M_0^2}{\sqrt{m^2
+\mathbf{k}^2_\perp}}\psi(x,{\bf{k}}_{\perp}),
\label{eq:17a}\\
\phi_{3}^\perp(x,\mu) &= \frac{\sqrt{2N_c}}{f_V}\int_0^{\mathbf{k}^2_\perp<\mu^2} \frac{d^2\mathbf{k}^2_\perp}{8\pi^3}\frac{\psi(x,{\bf{k}}_{\perp})}{\sqrt{m^2
+\mathbf{k}^2_\perp}}\frac{1}{M_0}[\frac{\mathbf{k}^2_\perp+m^2}{2x(1-x)}-\frac{M_0\mathbf{k}^2_\perp}{M_0+2m}],
\label{eq:17b}
\end{align}
where $\phi_{2(3)}$ is obtained by the $\mathbf{k}_\perp$ integration of the LF
wave function up to the transverse momentum scale $\mu$, $\mu$ can be regarded as the energy scale that separates
the perturbative and non-perturbative regimes. The normalization factor is $\mu_M= M^2/2m$. These DAs are usually expanded in terms of the Gegenbauer polynomials $C_n^{1/2}$ and $C_n^{3/2}$ as follows
\begin{align}
\phi_2^{A}(x,\mu)
&= \phi_{\rm as}(x)
   \sum_{n=0}^{\infty} a_n^{A}(\mu)\,
   C_n^{3/2}(2x-1),\\
\phi_3^{P}(x,\mu)
&= \phi_{3,\rm as}^{P}(x)\sum_{n=0}^{\infty} a_n^{P}(\mu)\,
   C_n^{1/2}(2x-1),\\
\phi_2^{\parallel}(x,\mu)
&= \phi_{\rm as}(x)
   \sum_{n=0}^{\infty} a_n^{\parallel}(\mu)\,
   C_n^{3/2}(2x-1),\\
\phi_3^{\perp}(x,\mu)
&= \phi^\perp_{3,\rm as}(x)
   \sum_{n=0}^{\infty} a_n^{\perp}(\mu)\,
   C_n^{3/2}(2x-1),
\label{eq:18}
\end{align}
where the asymptotic form LCDAs $\phi^A_{2,as}(x)=\phi^\parallel_{2,as}(x)=\phi_{as}(x)=6x\bar{x}$, $\phi_{3,\rm as}^{P}(x)=1$ and $\phi^\perp_{3,as}(x)=(3/4)(1+\xi^2)$. The explicit formula for $a_n(\mu)$ is given below
\begin{align}
a_n^{A}(\mu)
&= \frac{4n+6}{3n^2+9n+6}
   \int_0^1 dx\,
   C_n^{3/2}(2x-1)\,
   \phi_2^{A}(x,\mu),
 \\[0.2cm]
a_n^{P}(\mu)
&= (2n+1)
   \int_0^1 dx\,
   C_n^{1/2}(2x-1)\,
   \phi_3^{P}(x,\mu),
 \\[0.2cm]
a_n^{\parallel}(\mu)
&= \frac{4n+6}{3n^2+9n+6}
   \int_0^1 dx\,
   C_n^{3/2}(2x-1)\,
   \phi_2^{\parallel}(x,\mu),
\\[0.2cm]
a_n^{\perp}(\mu)
&= \frac{4n+6}{3n^2+9n+6}
   \int_0^1 dx\,
   C_n^{3/2}(2x-1)\,
   \phi_3^{\perp}(x,\mu).
\label{eq:19}
\end{align}
With the LCDAs of the meson, we can obtain additional information about the bound state's characteristics such as the $\xi$-moment
\begin{align}
\langle\xi^n\rangle^{A(P)} = \int_0^1 dx~\xi^n~\phi^{A(P)}_{2(3)}(x),\label{eq:20a}\\
\langle\xi^n\rangle^{\parallel(\perp)} = \int_0^1 dx~\xi^n~\phi^{\parallel(\perp)}_{2(3)}(x),
\label{eq:20}
\end{align}
where $\xi$ represents the longitudinal discrepancy between constituent quarks in the bound state. Similarly, the nonperturbative quantity of transverse momentum moments is also obtained by
\begin{align}
\langle\mathbf{k}^n_\perp\rangle^{A(P)} = \int_0^1 dx~\mathbf{k}^n_\perp~\phi^{A(P)}_{2(3)}(x),\label{eq:21a}\\
\langle\mathbf{k}^n_\perp\rangle^{\parallel(\perp)} = \int_0^1 dx~\mathbf{k}^n_\perp~\phi^{\parallel(\perp)}_{2(3)}(x),
\label{eq:21}
\end{align}
which reflects the information about the transverse size of the bound state.

\section{Numerical results} \label{sec:3}
\begin{table}[tbp]
\centering
\renewcommand{\tabcolsep}{0.3pc}
\caption{\label{tab:1}The inputs of  constituent quark masses [GeV] and Gaussian parameter $\beta$ [GeV] in our previous works~\cite{Xu:2025ntz,Chang:2019obq}.}
\begin{tabular}{ccccccccc}
				\hline\hline
   $m_q$ &
    $m_s$ & $m_c$ & $m_b$ & $\beta _{q\bar{q}}$ &
  $\beta _{s\bar{s}}$   & $\beta_{c\bar{c}}$ & $\beta_{b\bar{b}}$ \\
    \hline
  $0.25^{+0.01}_{-0.01}$ &
    $0.50^{+0.02}_{-0.02}$ & $1.80^{+0.09}_{-0.09}$ & $5.10^{+0.25}_{-0.25}$ & $0.321^{+0.016}_{-0.016}$ & $0.348^{+0.006}_{-0.006}$ & $0.703^{+0.007}_{-0.007}$ &
    $1.390^{+0.012}_{-0.012}$\\

		\hline
    \end{tabular}
\end{table}
In this section, we present the numerical results for the leading- and next-to-leading-twist LCDAs of pseudoscalar and vector quarkonium, together with their Gegenbauer moments, $\xi$-moments, and transverse momentum moments. Our analysis focuses on revealing the systematic evolution of longitudinal and transverse structures with increasing quark mass and on identifying universal features emerging in the heavy-quark limit.

The model parameters, including the constituent quark masses and the Gaussian parameter $\beta$, are listed in Table~\ref{tab:1}. These parameters are fixed by fitting the experimental values of meson decay constants~\cite{ParticleDataGroup:2022pth} and have been successfully used in our previous studies~\cite{Xu:2025ntz,Chang:2019obq}. The explicit formulae of the decay constants in LFQM are displayed as
\begin{eqnarray}
f_P = \frac{\sqrt{2N_c}}{8\pi^3}\int_0^1 dx \int d^2\mathbf{k}_\perp\frac{m}{\sqrt{\mathbf{k}^2_\perp+m^2}}\psi(x,\mathbf{k}_\perp)
\label{eq:22}
\end{eqnarray}
and
\begin{eqnarray}
f_V = \frac{\sqrt{2N_c}}{8\pi^3}\int_0^1 dx \int d^2\mathbf{k}_\perp\frac{m+\frac{2\mathbf{k}^2_\perp}{M_0+2m}}{\sqrt{\mathbf{k}^2_\perp+m^2}}
\psi(x,\mathbf{k}_\perp)
\label{eq:23}
\end{eqnarray}
for pseudoscalar quarkonium and vector quarkonium, respectively.
\subsection{Distribution amplitudes}\label{sec:3.1}

\begin{figure}[t]
\centering
\includegraphics[width=.45\textwidth]{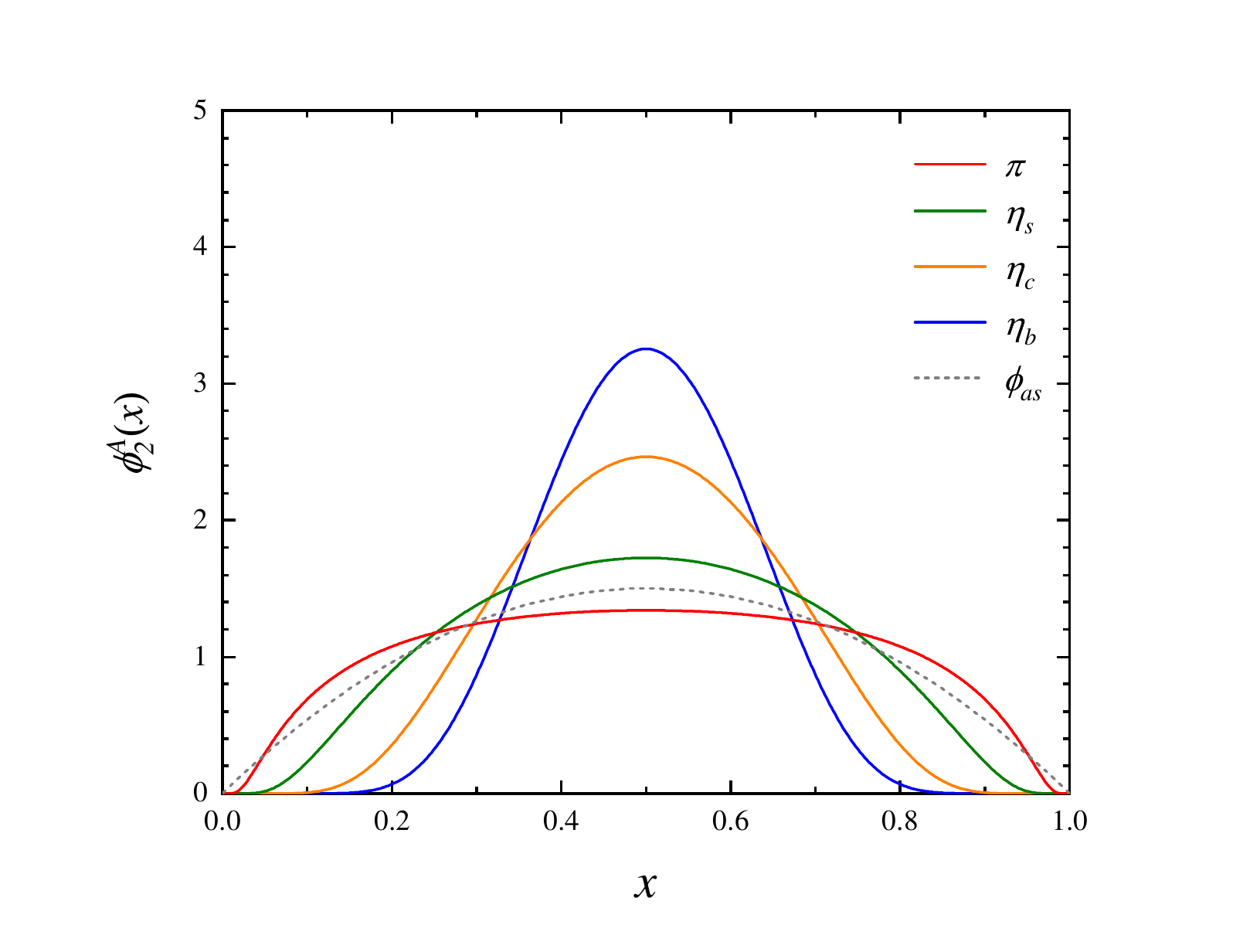}
\includegraphics[width=.45\textwidth]{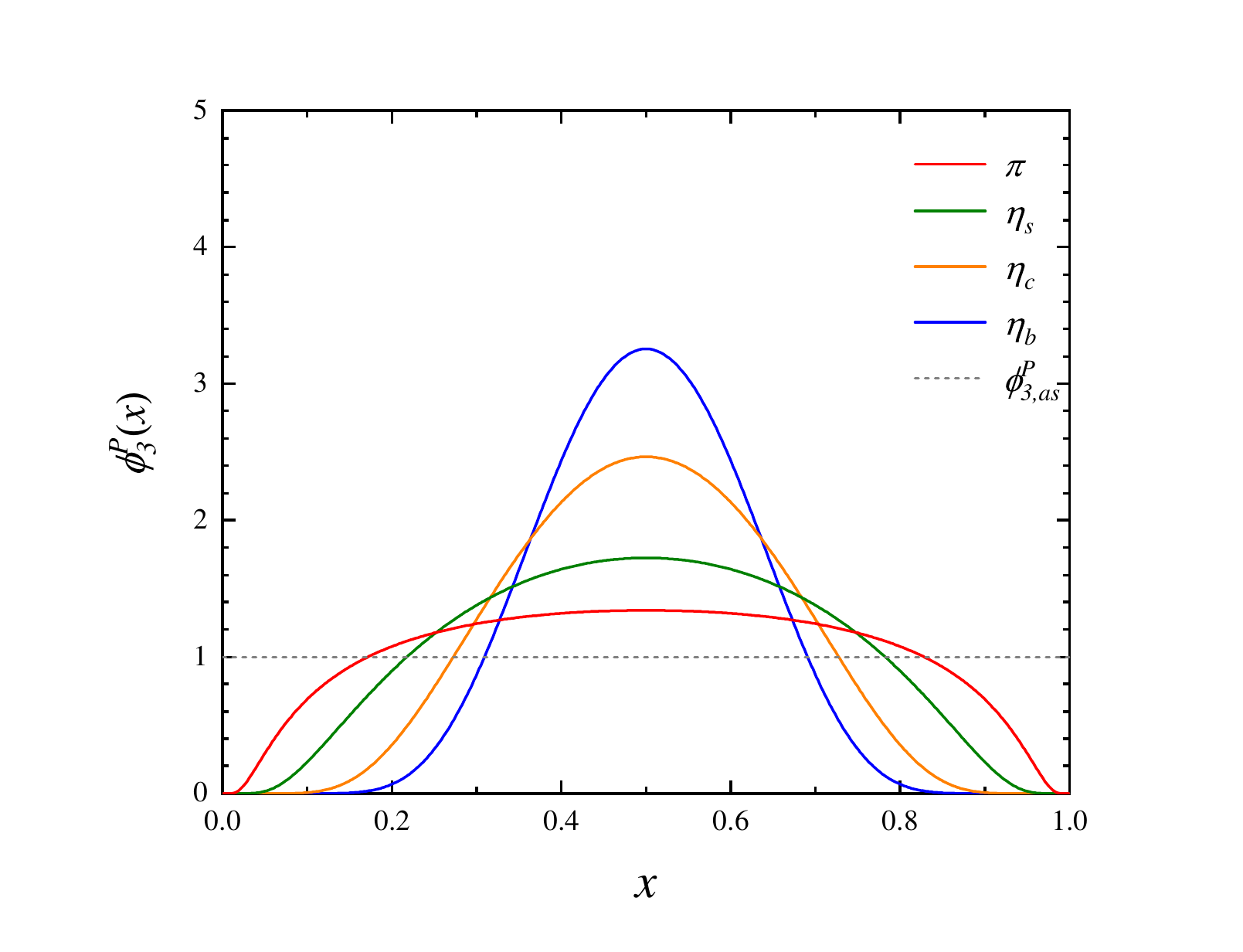}\\[2mm]
\includegraphics[width=.45\textwidth]{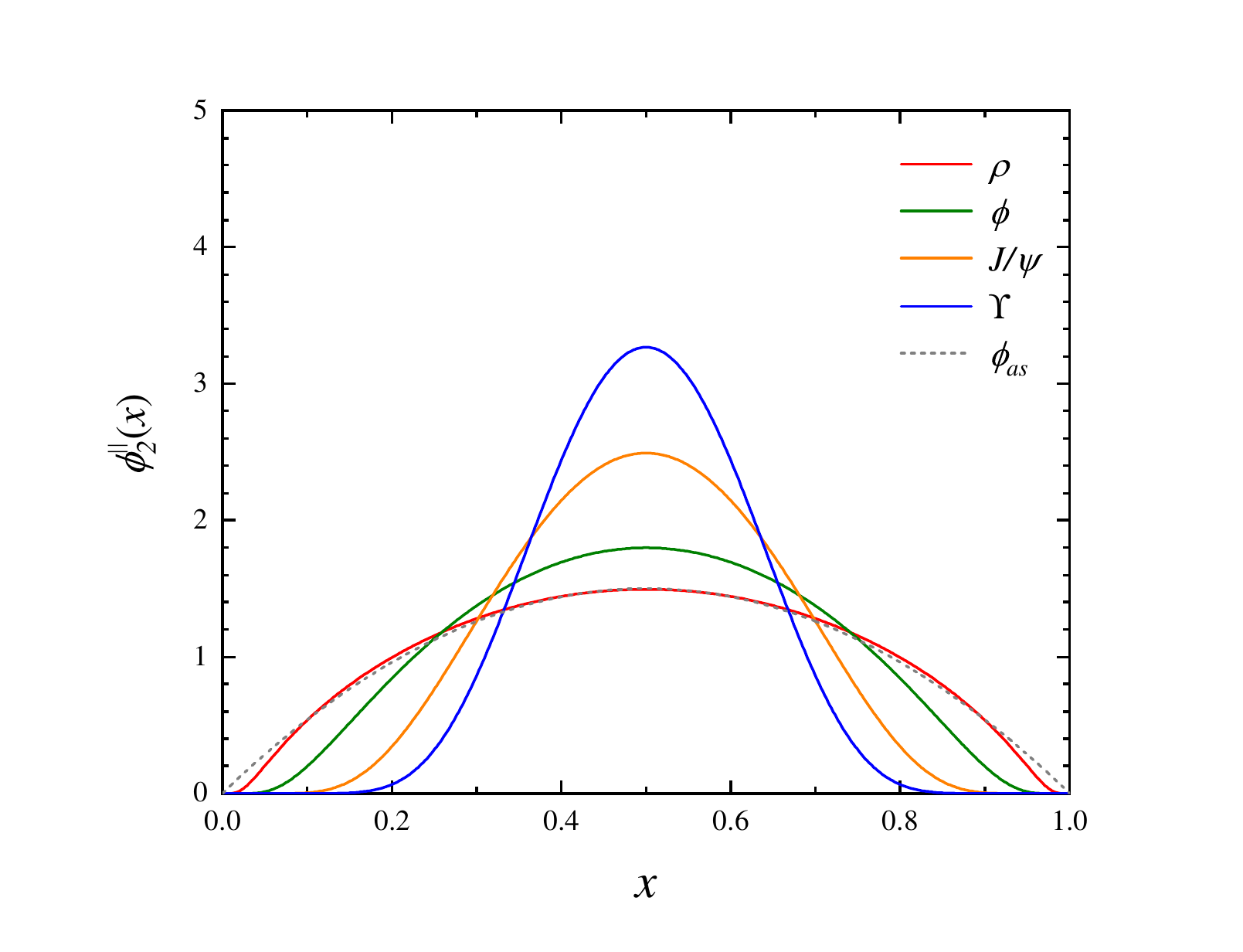}
\includegraphics[width=.45\textwidth]{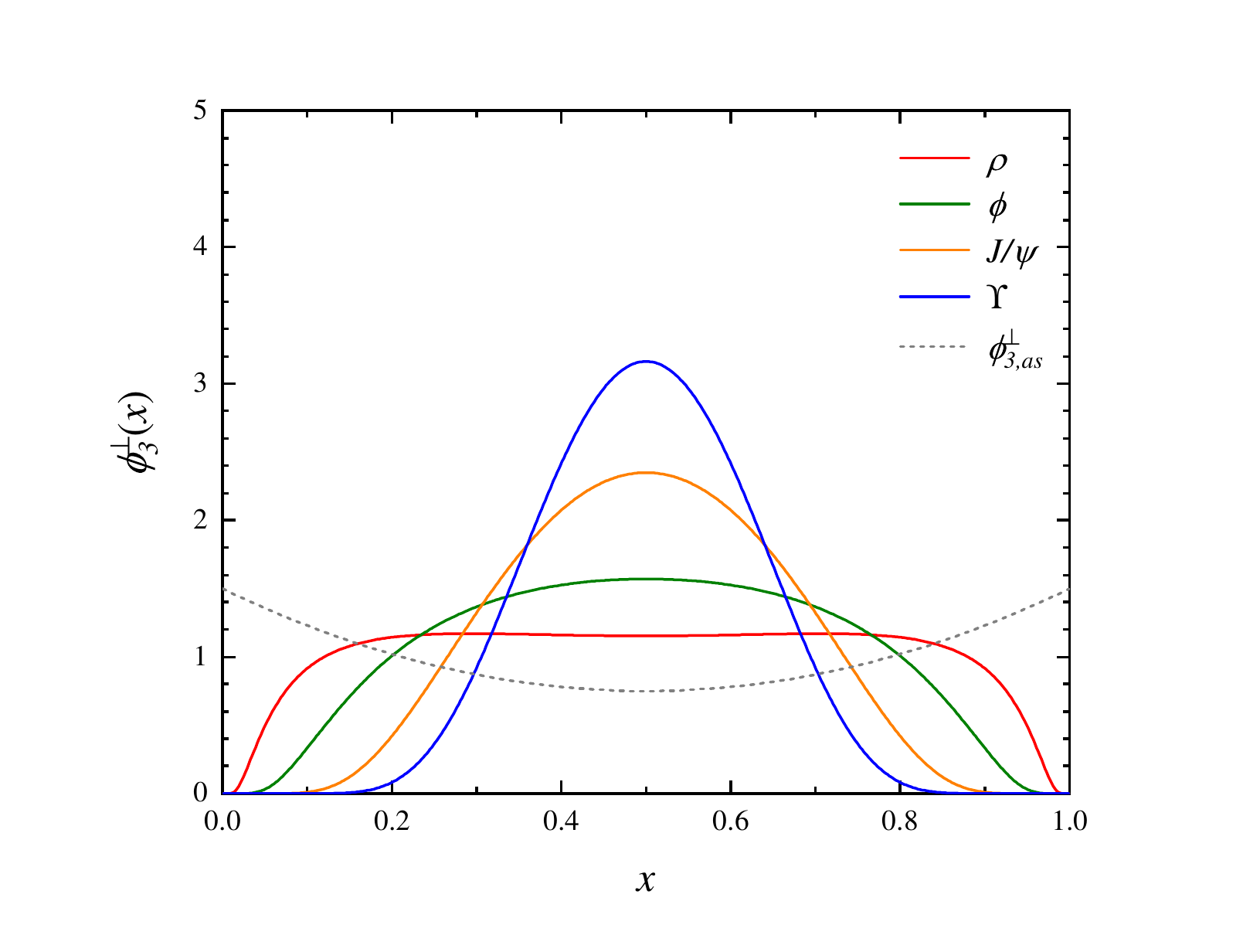}
\caption{The LCDAs of pseudoscalar quarkonium and their vector partners.}
\label{fig:1}
\end{figure}
The twist-2 and twist-3 LCDAs of pseudoscalar and vector quarkonium are shown in Figure~\ref{fig:1}. Several general and robust features can be identified.

First, owing to charge-conjugation symmetry, all quarkonium LCDAs are strictly symmetric under $x \leftrightarrow 1-x$. As a direct consequence, all odd Gegenbauer moments and odd $\xi$-moments vanish exactly. This symmetry property is manifest in the distributions displayed in Figure~\ref{fig:1} and provides an important consistency check of our numerical implementation.

Second, a clear and systematic mass dependence of the LCDAs is observed. As the constituent quark mass increases from light to heavy quarkonium, the distributions become increasingly peaked at $x=1/2$, while their widths around this point narrow significantly. This behavior indicates a progressive localization of the longitudinal momentum distribution and reflects the suppression of relativistic effects in heavy quark systems.

\begin{figure}[t]
\centering
\includegraphics[width=11cm,height=9.5cm]{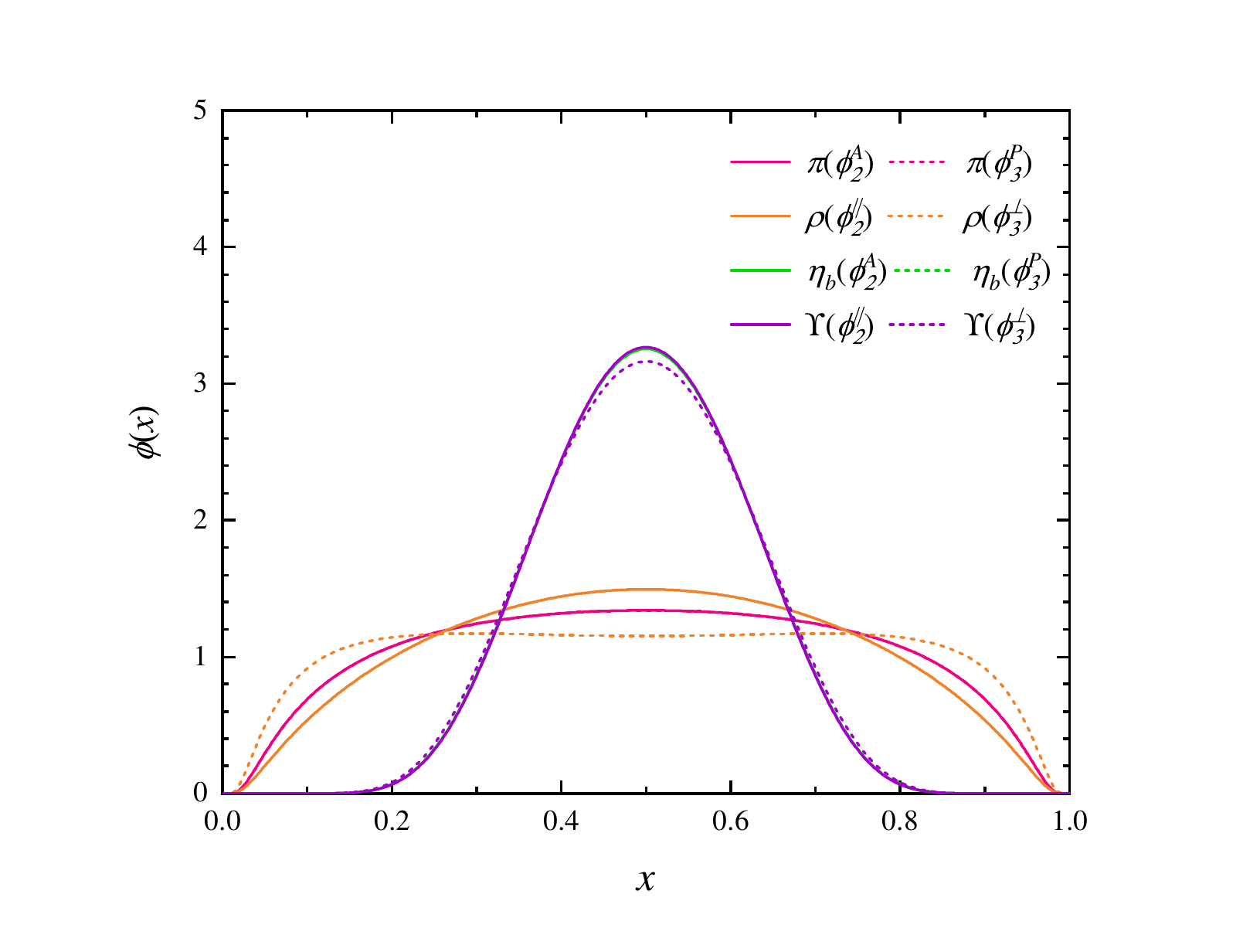}
\caption{The LCDAs of ($\pi$, $\rho$) and ($\eta_b, \Upsilon$).}
\label{fig:2}
\end{figure}

From Figure~\ref{fig:2}, a particularly notable feature emerges when comparing twist-2 and twist-3 LCDAs. For pseudoscalar quarkonium, the twist-2 and twist-3 distributions coincide exactly. This identity originates from the replacement $M \to M_0$, under which the analytic expressions (eqs.~(\ref{eq:16a},\ref{eq:17a})) for $\phi^A_{2}$ and $\phi^P_{3}$ become identical. Consequently, pseudoscalar quarkonium exhibits an exact twist-independence in both longitudinal and transverse momentum structures as implied in Table~\ref{tab:3} and~\ref{tab:4}.
We emphasize that the exact identity between twist-2 and twist-3 LCDAs for pseudoscalar quarkonium in the LFQM arises from the self-consistent replacement $M \to M_0$ and not a general prediction of QCD. Without this substitution, the two distributions are not identical. Generally, the heavy-quark limit corresponds to $m_q \to \infty$ Within the LFQM framework, this limit can be quantitatively characterized by the ratio $m/\beta$, which measures the degree of nonrelativistic behavior of the system. As shown in Table~\ref{tab:1}, $m/\beta$ increases significantly from light quarks ($u,d$) to heavy quarks ($c,b$), reaching its largest value in bottomonium, which lies closest to the heavy-quark limit in this framework. Figures~\ref{fig:1} and~\ref{fig:2} further show that for $\eta_c$, $J/\psi$, $\eta_b$, and $\Upsilon$, the twist-2 and twist-3 LCDAs approach each other as $m/\beta$ becomes large, providing a quantitative criterion for the heavy quark regime in this model. It is interesting to note that using our fitted parameters, we obtain $\beta/m \approx 0.39$ for $J/\psi$ and $\beta/m \approx 0.24$ for $\Upsilon(1S)$. Meanwhile, a Bethe--Salpeter study~\cite{Wang:2020zbr} gives the velocity estimates $v \approx 0.46$ for $J/\psi$ and $v \approx 0.26$ for $\Upsilon(1S)$. This suggests that the parameter ratio $\beta/m$ in the LFQM may be related in some way to the velocity parameter $v$. In this sense, two levels can be distinguished: the exact twist-independence in pseudoscalar quarkonium arises from the $M \to M_0$ prescription within the self-consistent LFQM, whereas the gradual convergence of twist-2 and twist-3 LCDAs with increasing quark mass reflects a more general dynamical suppression of relativistic effects. In the limit $m_q \to \infty$, even without this replacement, the two distributions naturally coincide.

\begin{figure}
\centering
\includegraphics[width=11cm,height=9.5cm]{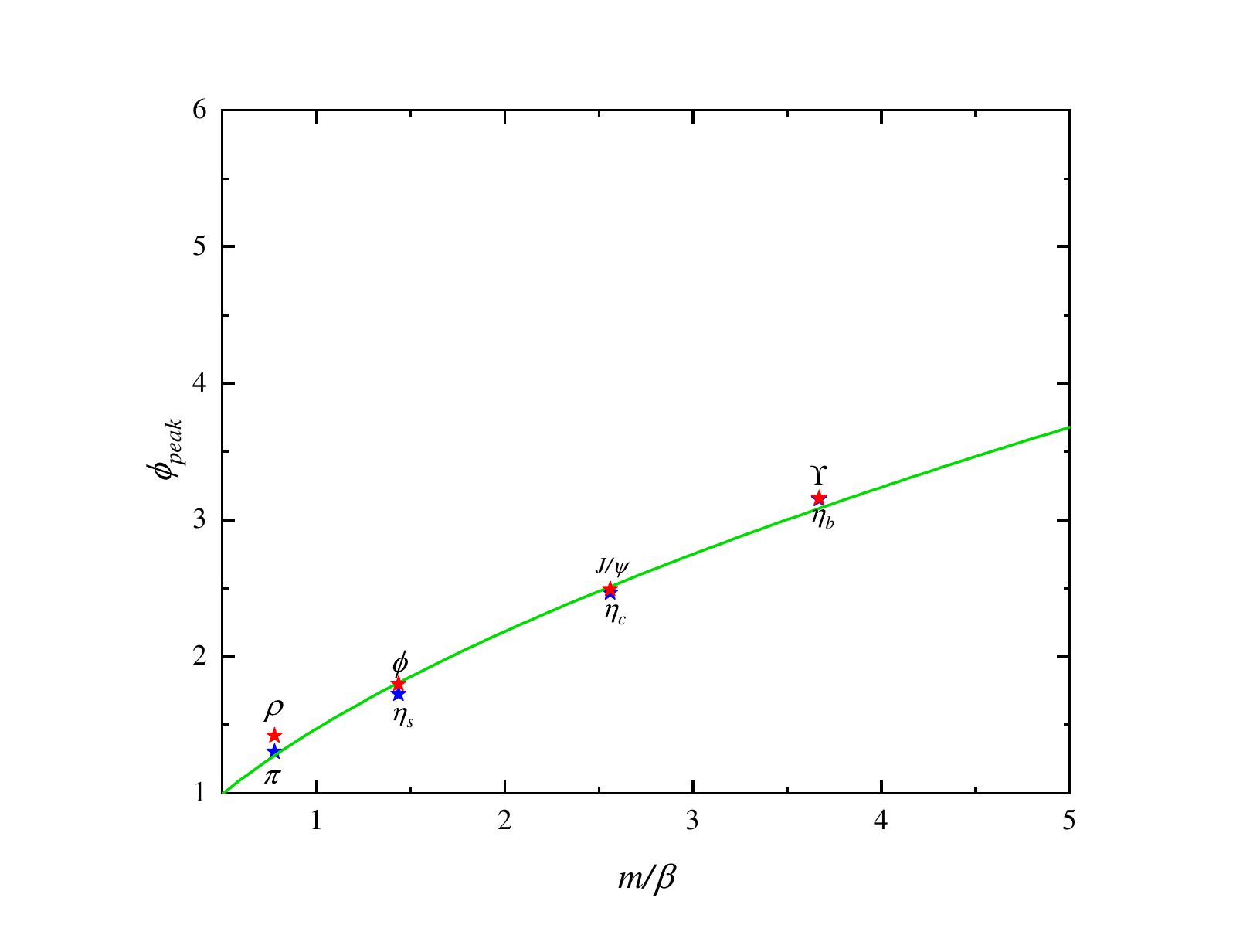}
\caption{The relationship between $m/\beta$ and peak value of LCDAs.}
\label{fig:3}
\end{figure}

For vector quarkonium, the twist-2 and twist-3 LCDAs are distinct at finite quark mass, with the twist-3 distribution exhibiting a slightly higher central peak. However, as the quark mass increases, the difference between the two twists diminishes rapidly. In the heavy-quark limit, eqs.~(\ref{eq:21a},\ref{eq:21}) and eqs.~(\ref{eq:16a}-\ref{eq:17b}) show that the decay constants satisfy $f_V \simeq f_P$, which in turn implies $\phi_2^{A} \simeq \phi_2^{\parallel}$. Consequently, the twist-2 and twist-3 LCDAs converge, $\phi_2^{\parallel} \simeq \phi_3^{\perp}$. Together with the pseudoscalar case, this leads to $\phi_2^{A} = \phi_3^{P} \simeq \phi_2^{\parallel} \simeq \phi_3^{\perp}$, demonstrating an emergent twist-independence of quarkonium LCDAs. Notably, this behavior should be viewed as an approximate numerical artifact of the model, not a strict theoretical consequence. Although this relation is formally model-dependent, similar tendencies can be found in Refs.~\cite{Hwang:2008qi,Choi:2013mda,Arifi:2022pal,Li:2017mlw,Zhong:2014fma,Zhong:2016kuv}, suggesting that twist-independence may represent a robust feature of heavy quark dynamics.

Furthermore, by analyzing the peak height of the pseudoscalar quarkonium LCDAs shown in Figure~\ref{fig:1}, we find a simple phenomenological scaling behavior. Within our model and parameter set, the peak height can be parametrized as $\phi_{\text{peak}} \simeq 1.47\,(m/\beta)^{0.57}$, Figure~\ref{fig:3} illustrating the phenomenological scaling behavior of $\phi_{\text{peak}}$. Numerically, this convergence pattern becomes already significant at the charmonium scale and is
further strengthened for bottomonium, indicating that the emergent twist-independence is not merely
a formal asymptotic feature but has phenomenological relevance.

\subsection{Gegenbauer moments $a_{n}$}\label{sec:3.2}
\begin{table}[htbp]
\caption{The Gegenbauer moments of twist-2 and twist-3 quarkonium LCDAs.}
\centering
\renewcommand{\arraystretch}{1.3}
\renewcommand{\tabcolsep}{1.2pc}
\begin{tabular}{c c c c c c}
\hline\hline
   P-meson  & $a_2^A$ & $a_4^A$& & $a_2^P$ & $a_4^P$ \\
\hline
$\pi$     & $0.054^{+0.024}_{-0.028}$ & $-0.034^{+0.010}_{-0.007}$    & &$-0.864^{+0.068}_{-0.069}$ & $-0.273^{+0.038}_{-0.028}$   \\
$\eta_s$   &  $-0.147^{+0.020}_{-0.020}$ & $-0.038^{+0.007}_{-0.005}$  & &$-1.378^{+0.051}_{-0.041}$ & $0.223^{+0.077}_{-0.073}$ \\
$\eta_c$     & $-0.338^{+0.019}_{-0.017}$ & $0.081^{+0.018}_{-0.017}$ & &$-1.869^{+0.048}_{-0.042}$ & $1.202^{+0.104}_{-0.118}$ \\
$\eta_b$      & $-0.431^{+0.013}_{-0.012}$ & $0.191^{+0.017}_{-0.018}$ & &$-2.109^{+0.035}_{-0.031}$ & $1.890^{+0.100}_{-0.110}$ \\
\hline
   V-meson  & $a_2^\parallel$ & $a_4^\parallel$& & $a_2^\perp$ & $a_4^\perp$ \\
\hline
$\rho$     & $-0.019^{+0.015}_{-0.017}$ & $-0.030^{+0.005}_{-0.005}$    & &$0.159^{+0.033}_{-0.034}$ & $-0.017^{+0.016}_{-0.017}$   \\
$\phi$       & $-0.168^{+0.017}_{-0.017}$ & $-0.026^{+0.005}_{-0.005}$ & &$-0.093^{+0.030}_{-0.024}$ & $-0.058^{+0.005}_{-0.004}$ \\
$J/\psi$     & $-0.342^{+0.018}_{-0.016}$ & $0.085^{+0.017}_{-0.016}$ & &$-0.320^{+0.022}_{-0.019}$ & $0.063^{+0.019}_{-0.019}$ \\
$\Upsilon$      & $-0.432^{+0.014}_{-0.012}$ & $0.192^{+0.017}_{-0.018}$ & &$-0.424^{+0.015}_{-0.013}$ & $0.181^{+0.018}_{-0.020}$ \\
\hline
\end{tabular}

\label{tab:2}
\end{table}
The Gegenbauer moments of twist-2 and twist-3 quarkonium LCDAs are summarized in Table~\ref{tab:2}. As required by charge-conjugation symmetry, all odd Gegenbauer moments vanish exactly. The even moments exhibit a clear and systematic evolution with quark mass: the second moments $a^A_2$ and $a_2^\perp$ decrease and becomes increasingly negative, while the fourth moment $a_4$ changes sign and grows in magnitude. These trends quantitatively reflect the sharpening of the DA peak and the narrowing of its width around $x=1/2$.

For light mesons with identical quark content, noticeable differences exist between pseudoscalar and vector states. In contrast, for heavy quarkonium partners such as $(\eta_c,J/\psi)$ and $(\eta_b,\Upsilon)$, these differences are strongly suppressed, providing further evidence for the universality of quarkonium LCDAs in the heavy-quark limit.

\subsection{$\xi$-moments $\langle\xi^n\rangle$}\label{sec:3.3}
\begin{table}[htb]
\caption{The $\xi$-moments of twist-2 and twist-3 quarkonium LCDAs.}
\centering
\renewcommand{\arraystretch}{1.2}
\renewcommand{\tabcolsep}{0.5pc}
\begin{tabular}{c c c c c c c c}
\hline\hline
\multirow{2}{*}{P-meson} & &{$\langle\xi^n\rangle^A$ } & & & &{$\langle\xi^n\rangle^P$}&\\
\cline{2-4}\cline{6-8}
     & $\langle\xi^2\rangle$ & $\langle\xi^4\rangle$& $\langle\xi^6\rangle$& & $\langle\xi^2\rangle$ & $\langle\xi^4\rangle$& $\langle\xi^6\rangle$ \\
\hline
$\pi$     & $0.218^{+0.009}_{-0.009}$ & $0.094^{+0.007}_{-0.007}$   &$0.051^{+0.006}_{-0.005}$ & &$0.218^{+0.009}_{-0.009}$ & $0.094^{+0.007}_{-0.007}$   &$0.051^{+0.006}_{-0.005}$   \\
$\eta_s$       & $0.149^{+0.007}_{-0.006}$ & $0.048^{+0.004}_{-0.004}$& $0.020^{+0.003}_{-0.002}$ & &$0.149^{+0.007}_{-0.006}$ & $0.048^{+0.004}_{-0.004}$& $0.020^{+0.003}_{-0.002}$ \\
$\eta_c$     & $0.084^{+0.006}_{-0.006}$ & $0.017^{+0.002}_{-0.002}$ &$0.005^{+0.001}_{-0.001}$ & &$0.084^{+0.006}_{-0.006}$ & $0.017^{+0.002}_{-0.002}$ &$0.005^{+0.001}_{-0.001}$ \\
$\eta_b$      & $0.052^{+0.005}_{-0.004}$ & $0.007^{+0.001}_{-0.001}$ &$0.001^{+0.001}_{-0.000}$ & &$0.052^{+0.005}_{-0.004}$ & $0.007^{+0.001}_{-0.001}$ &$0.001^{+0.001}_{-0.000}$\\
\hline
\multirow{2}{*}{V-meson} & &{$\langle\xi^n\rangle^\parallel$} & & & &{$\langle\xi^n\rangle^\perp$}&\\
\cline{2-4}\cline{6-8}
     & $\langle\xi^2\rangle$ & $\langle\xi^4\rangle$& $\langle\xi^6\rangle$& & $\langle\xi^2\rangle$ & $\langle\xi^4\rangle$& $\langle\xi^6\rangle$ \\
\hline
$\rho$     & $0.193^{+0.006}_{-0.006}$ & $0.078^{+0.004}_{-0.004}$   &$0.040^{+0.003}_{-0.003}$ & &$0.255^{+0.011}_{-0.012}$ & $0.120^{+0.009}_{-0.009}$& $0.069^{+0.008}_{-0.007}$   \\
$\phi$       & $0.142^{+0.006}_{-0.006}$ & $0.044^{+0.004}_{-0.003}$ &$0.018^{+0.002}_{-0.002}$ & &$0.168^{+0.008}_{-0.008}$ & $0.058^{+0.005}_{-0.005}$& $0.026^{+0.003}_{-0.003}$ \\
$J/\psi$     & $0.083^{+0.006}_{-0.006}$ & $0.016^{+0.003}_{-0.002}$ &$0.005^{+0.001}_{-0.001}$ & &$0.090^{+0.008}_{-0.006}$ & $0.019^{+0.003}_{-0.002}$& $0.006^{+0.001}_{-0.001}$ \\
$\Upsilon$      & $0.052^{+0.004}_{-0.005}$ & $0.007^{+0.001}_{-0.001}$ &$0.001^{+0.001}_{-0.000}$ & &$0.055^{+0.005}_{-0.005}$ & $0.008^{+0.001}_{-0.002}$& $0.001^{+0.001}_{-0.000}$ \\
\hline
\end{tabular}

\label{tab:3}
\end{table}

\begin{table}[htb]
\caption{Comparison of $\xi$-moments results in heavy quarkonium with different theoretical models.}
\centering
\small
\renewcommand{\arraystretch}{1.2}
\renewcommand{\tabcolsep}{0.03pc}
\begin{tabular}{ c c c c c c c c c c}
\hline\hline
         $\langle\xi^n\rangle$&This work & NRQCD\cite{Bodwin:2006dn} & QCDSR\cite{Braguta:2006wr,Braguta:2007fh} & DSE\cite{Ding:2015rkn} & LFQM\cite{Arifi:2022pal} &BLFQ\cite{Li:2017mlw} & pQCD\cite{Lepage:1980fj} & AdS/QCD\cite{Brodsky:2014yha}\\
    \hline
        $\langle\xi^2\rangle^A_{\eta_c}$ &$0.084^{+0.006}_{-0.006}$ & $0.075^{+0.011}_{-0.011}$ & $0.070^{+0.007}_{-0.007}$ &$0.100$ & 0.088& $0.096^{+0.013}_{-0.013}$ & 0.200& 0.250\\
$\langle\xi^4\rangle^A_{\eta_c}$ &$0.017^{+0.002}_{-0.002}$ & $0.010^{+0.003}_{-0.003}$ & $0.012^{+0.002}_{-0.002}$ &$0.032$& 0.018 & $0.019^{+0.002}_{-0.002}$ & 0.086&0.130 \\
 $\langle\xi^6\rangle^A_{\eta_c}$&$0.005^{+0.001}_{-0.001}$ & $0.0017^{+0.0007}_{-0.0007}$ & $0.0032^{+0.0009}_{-0.0009}$ &$ 0.015$& 0.005 & $0.0036^{+0.0027}_{-0.0027}$ &0.047& 0.078\\
  $\mu$&$\sim$2 GeV & $m_c$ & $m_c$ &2 GeV& $\sim$2 GeV & $m_c$ & $\infty$&-- \\
\hline
 $\langle\xi^2\rangle^\parallel_{J/\psi}$&$0.083^{+0.006}_{-0.006}$ & $0.075^{+0.011}_{-0.011}$ & $0.070^{+0.007}_{-0.007}$ &$ 0.039$& 0.086 & $0.096^{+0.020}_{-0.020}$ & 0.200& 0.250\\
     $\langle\xi^4\rangle^\parallel_{J/\psi}$ &$0.016^{+0.003}_{-0.002}$& $0.010^{+0.003}_{-0.003}$ & $0.012^{+0.002}_{-0.002}$ & $0.0038$& 0.017 & $0.021^{+0.009}_{-0.009}$ & 0.086 & 0.130 \\
$\langle\xi^6\rangle^\parallel_{J/\psi}$&$0.005^{+0.001}_{-0.001}$ & $0.0017^{+0.0007}_{-0.0007}$ & $0.0032^{+0.0009}_{-0.0009}$ &$ 7.3\times10^{-4}$& 0.005 & $0.0060^{+0.0041}_{-0.0041}$ & 0.047& 0.078\\
  $\mu$&$\sim$2 GeV & $m_c$ & $m_c$ &2 GeV& $\sim$2 GeV & $m_c$ & $\infty$&-- \\
 \hline

 $\langle\xi^2\rangle^A_{\eta_b}$&$0.052^{+0.005}_{-0.004}$ & -- & -- &$ 0.070$& 0.049 & $0.052^{+0.002}_{-0.002}$ & 0.200&0.250 \\
     $\langle\xi^4\rangle^A_{\eta_b}$ &$0.007^{+0.001}_{-0.001}$& -- & -- & 0.015& 0.006  & $0.0081^{+0.0061}_{-0.0061}$ & 0.086 &0.130\\
          $\langle\xi^6\rangle^A_{\eta_b}$ &$0.001^{+0.001}_{-0.000}$ & -- & -- &$ 0.0042$& 0.001 & $0.0020^{+0.0048}_{-0.0048}$ & 0.047&0.078 \\
            $\mu$&$\sim$4 GeV & $m_b$ & $m_b$ &2 GeV& $\sim$4 GeV & $m_b$ & $\infty$&-- \\
          \hline

           $\langle\xi^2\rangle^\parallel_{\Upsilon}$&$0.052^{+0.004}_{-0.005}$ & -- & -- &$ 0.014$& 0.049 & $0.047^{+0.017}_{-0.017}$ & 0.200& 0.250\\
     $\langle\xi^4\rangle^\parallel_{\Upsilon}$ & $0.007^{+0.001}_{-0.001}$ & -- & -- & $ 4.3\times10^{-4}$& 0.006  &$0.0066^{+0.0073}_{-0.0073}$ &0.086 &0.130\\
          $\langle\xi^6\rangle^\parallel_{\Upsilon}$ & $0.001^{+0.001}_{-0.000}$ & -- & -- &$ 4.4\times10^{-5}$& 0.001  & $0.0014^{+0.0063}_{-0.0063}$ & 0.047&0.078 \\
            $\mu$&$\sim$4 GeV & $m_b$ & $m_b$ &2 GeV& $\sim$4 GeV & $m_b$ & $\infty$&-- \\
\hline
\end{tabular}

\label{tab:4}
\end{table}
The $\xi$-moments of twist-2 and twist-3 quarkonium LCDAs are presented in Table~\ref{tab:3}. For pseudoscalar quarkonium, the identity $\phi^A_{2}=\phi^P_{3}$ implies identical $\xi$-moments at different twists. The $\xi$-moments decrease rapidly with increasing moment order and with increasing quark mass, indicating strong suppression of endpoint contributions and a highly localized longitudinal momentum distribution. For vector quarkonium, the $\xi$-moments are twist-dependent at finite quark mass, but this dependence becomes progressively weaker as the quark mass increases. For charmonium and bottomonium, the twist-2 and twist-3 $\xi$-moments are already very close, consistent with the convergence of LCDAs in the heavy-quark limit.

Table~\ref{tab:4} presents a comparison of the first few moments of selected heavy quarkonium states obtained from various theoretical approaches. For a systematic comparison, results from several representative frameworks are included, namely NRQCD~\cite{Bodwin:2006dn}, QCDSR~\cite{Braguta:2006wr,Braguta:2007fh}, the DSE~\cite{Ding:2015rkn}, the LFQM~\cite{Arifi:2022pal}, and Basis Light-Front Quantization (BLFQ~\cite{Li:2017mlw}).
In most of these approaches, the moments are evaluated at the effective heavy quark mass scale, $\mu \approx m_q$, except for the DSE results, which are given at $\mu = 2~\text{GeV}$. For completeness, the pQCD asymptotic results~\cite{Lepage:1980fj} and the AdS/QCD model proposed by Brodsky and de T\'eramond (AdS/QCD~\cite{Brodsky:2014yha}) are also listed. However, these two approaches are not strictly applicable to heavy quarkonium at the heavy quark mass scale. Earlier LFQM~\cite{Arifi:2022pal} observe that $\mathbf{k}_\perp\to \infty$ corresponds to energy scales $\mu$ of approximately 2~GeV for $\eta_c$ and 4~GeV for $\eta_b$, respectively. Since the primary objective of this work is to compare different twist structures and meson systems at a common effective scale, without loss of generality we adopt the same treatment as in Ref.~\cite{Arifi:2022pal} to evaluate the LCDAs. This procedure ensures the proper normalization of the LCDAs, which does not qualitatively affect the patterns observed in our analysis.
\begin{table}[htbp]
\caption{The transverse momentum moments $\sqrt[n]{\langle\mathbf{k}_\perp^n\rangle}$ of twist-2 and twist-3 quarkonium LCDAs.}
\centering
\small
\renewcommand{\arraystretch}{1.3}
\renewcommand{\tabcolsep}{0.1pc}
\begin{tabular}{c c c c c c c c c c}
\hline\hline
\multirow{2}{*}{P-meson} &&\multicolumn{2}{c}{$\sqrt[n]{\langle\mathbf{k}_\perp^n\rangle}^A$ } & &&&\multicolumn{2}{c}{$\sqrt[n]{\langle\mathbf{k}_\perp^n\rangle}^P$ }&\\
\cline{2-5}\cline{7-10}
     & $\langle\mathbf{k}_\perp\rangle$ & $\sqrt{\langle \mathbf{k}^2_\perp\rangle}$& $\sqrt[3]{\langle \mathbf{k}^3_\perp\rangle}$& $\sqrt[4]{\langle \mathbf{k}^4_\perp\rangle}$& & $\langle\mathbf{k}_\perp\rangle$ & $\sqrt{\langle \mathbf{k}^2_\perp\rangle}$& $\sqrt[3]{\langle \mathbf{k}^3_\perp\rangle}$& $\sqrt[4]{\langle \mathbf{k}^4_\perp\rangle}$ \\
\hline
$\pi$       & $0.323^{+0.012}_{-0.012}$ & $0.372^{+0.015}_{-0.015}$   &$0.416^{+0.017}_{-0.017}$ &$0.456^{+0.020}_{-0.020}$ & &$0.323^{+0.012}_{-0.012}$ & $0.372^{+0.015}_{-0.015}$   &$0.416^{+0.017}_{-0.017}$ &$0.456^{+0.020}_{-0.020}$   \\
$\eta_s$       & $0.381^{+0.003}_{-0.004}$ & $0.433^{+0.005}_{-0.004}$ &$0.480^{+0.005}_{-0.006}$ &$0.522^{+0.007}_{-0.006}$ & &$0.381^{+0.003}_{-0.004}$ & $0.433^{+0.005}_{-0.004}$ &$0.480^{+0.005}_{-0.006}$ &$0.522^{+0.007}_{-0.006}$ \\
$\eta_c$     & $0.821^{+0.004}_{-0.003}$ & $0.929^{+0.004}_{-0.004}$ &$1.024^{+0.005}_{-0.005}$ &$1.110^{+0.006}_{-0.006}$ & &$0.821^{+0.004}_{-0.003}$ & $0.929^{+0.004}_{-0.004}$ &$1.024^{+0.005}_{-0.005}$ &$1.110^{+0.006}_{-0.006}$ \\
$\eta_b$   & $1.671^{+0.008}_{-0.009}$ & $1.887^{+0.009}_{-0.010}$ &$2.077^{+0.010}_{-0.011}$ &$2.248^{+0.012}_{-0.012}$ & &$1.671^{+0.008}_{-0.009}$ & $1.887^{+0.009}_{-0.010}$ &$2.077^{+0.010}_{-0.011}$ &$2.248^{+0.012}_{-0.012}$ \\
\hline
\multirow{2}{*}{V-meson} &&\multicolumn{2}{c}{$\sqrt[n]{\langle\mathbf{k}_\perp^n\rangle}^\parallel$} & &&&\multicolumn{2}{c}{$\sqrt[n]{\langle\mathbf{k}_\perp^n\rangle}^\perp$}&\\
\cline{2-5}\cline{7-10}
     & $\langle\mathbf{k}_\perp\rangle$ & $\sqrt{\langle \mathbf{k}^2_\perp\rangle}$& $\sqrt[3]{\langle \mathbf{k}^3_\perp\rangle}$& $\sqrt[4]{\langle \mathbf{k}^4_\perp\rangle}$ & & $\langle\mathbf{k}_\perp\rangle$&$\sqrt{\langle \mathbf{k}^2_\perp\rangle}$& $\sqrt[3]{\langle \mathbf{k}^3_\perp\rangle}$& $\sqrt[4]{\langle \mathbf{k}^4_\perp\rangle}$ \\
\hline
$\rho$       & $0.385^{+0.020}_{-0.019}$ & $0.436^{+0.022}_{-0.021}$   &$0.481^{+0.024}_{-0.024}$ &$0.521^{+0.027}_{-0.025}$ & &$0.354^{+0.016}_{-0.016}$ & $0.405^{+0.018}_{-0.019}$& $0.450^{+0.021}_{-0.021}$&$0.491^{+0.023}_{-0.023}$   \\
$\phi$       & $0.421^{+0.007}_{-0.007}$ & $0.476^{+0.008}_{-0.007}$ &$0.525^{+0.008}_{-0.009}$ &$0.568^{+0.010}_{-0.011}$ & &$0.401^{+0.005}_{-0.006}$ & $0.455^{+0.006}_{-0.007}$& $0.503^{+0.007}_{-0.008}$&$0.546^{+0.008}_{-0.008}$ \\
$J/\psi$     & $0.863^{+0.007}_{-0.008}$ & $0.974^{+0.009}_{-0.008}$ &$1.072^{+0.009}_{-0.009}$ &$1.160^{+0.010}_{-0.010}$ & &$0.841^{+0.006}_{-0.005}$ & $0.951^{+0.007}_{-0.006}$& $1.048^{+0.007}_{-0.007}$&$1.135^{+0.008}_{-0.008}$ \\
$\Upsilon$   & $1.720^{+0.012}_{-0.014}$ & $1.941^{+0.014}_{-0.015}$ &$2.135^{+0.016}_{-0.017}$ &$2.309^{+0.018}_{-0.018}$ & &$1.695^{+0.010}_{-0.011}$ & $1.914^{+0.011}_{-0.013}$& $2.106^{+0.013}_{-0.014}$&$2.279^{+0.014}_{-0.015}$ \\
\hline
\end{tabular}

\label{tab:5}
\end{table}
\subsection{Transverse momentum moments $\sqrt[n]{\langle\mathbf{k}_\perp^n\rangle}$}\label{sec:3.4}


The transverse momentum moments $\sqrt[n]{\langle \mathbf{k}_\perp^n \rangle}$ are listed in Table~\ref{tab:5}. For pseudoscalar quarkonium, twist-2 and twist-3 transverse moments are identical,  reflecting exact twist independence. The transverse moments increase with both the moment order and the quarkonium mass, indicating a larger characteristic transverse momentum scale and a more compact spatial structure. Interestingly, this observation supports our conclusion in recent work~\cite{Xu:2025ntz} that the charge radii of mesons decrease gradually with increasing constituent quark mass.

For vector quarkonium, a similar trend is observed. The transverse moments are slightly larger than those of pseudoscalar states with the same quark content, suggesting a broader transverse momentum distribution. As in the longitudinal case, the difference between twist-2 and twist-3 transverse moments becomes increasingly small for heavy quarkonium, reinforcing the picture of emergent twist-independence.
\section{Summary}\label{sec:4}

In this work, we systematically investigate the twist-2 and twist-3 LCDAs of ground-state quarkonium within the LFQM. We present a comprehensive set of pseudoscalar and vector LCDAs and compute several related observables, including Gegenbauer moments, $\xi$-moments, and transverse momentum moments, together with analytical derivations and numerical cross-checks. We further identify an approximate scaling relation between the LCDA peak value and the model parameters. The transverse momentum moments are found to increase with meson mass, suggesting a more compact bound-state configuration. In the heavy-quark limit, the LCDAs exhibit convergence toward twist-independence, consistent with NRQCD expectations that higher-twist and spin-dependent effects are parametrically suppressed. With the replacement $M \to M_0$, we consistently analyze both longitudinal and transverse momentum structures. Overall, our results extend previous LFQM studies by providing a unified and systematic analysis of different twist structures in heavy quarkonium.


Owing to charge-conjugation symmetry, the LCDAs of all quarkonium states are symmetric under $x \leftrightarrow 1-x$, leading to the exact vanishing of all odd-order Gegenbauer moments and odd $\xi$-moments. We find that the quark mass plays a decisive role in shaping the LCDAs. As the constituent quark mass increases, the distribution amplitudes become increasingly peaked at $x=1/2$ and their widths narrow significantly, signaling a highly localized longitudinal momentum distribution and the suppression of relativistic effects in heavy quarkonium.

A central result of this study is the emergence of an approximate \emph{twist-independence} within the LFQM framework. For pseudoscalar quarkonium, the replacement $M \to M_0$ leads to an exact identity between the twist-2 and twist-3 distribution amplitudes, implying identical longitudinal and transverse momentum structures. It should be reiterated that this equality is a model-dependent feature of the self-consistent LFQM formulation and not a general prediction of QCD. For vector quarkonium, while twist-2 and twist-3 LCDAs differ at finite quark mass, they gradually converge as the quark mass increases. In the heavy-quark limit, $\phi^A_{2}\simeq \phi^{\parallel}_{2}$, then all quarkonium LCDAs satisfy
$\phi^A_{2} = \phi^P_{3} \simeq \phi^{\parallel}_{2} \simeq \phi^{\perp}_{3}$,
indicating that twist distinctions become physically irrelevant and that the dynamics is governed by a universal nonrelativistic structure. We emphasize that this behavior is an approximate numerical trend observed
within the self-consistent LFQM, rather than a strict theoretical prediction.

We have also shown that the transverse momentum moments increase with both the moment order and the meson mass, reflecting a progressively more compact spatial structure of heavy quarkonium. This behavior is consistent with the Heisenberg's uncertainty principle and with previous findings on the mass dependence of meson charge radii. Furthermore, for pseudoscalar quarkonium, we identify a simple phenomenological scaling law for the peak value of the distribution amplitude, controlled primarily by the ratio $m/\beta$ within our model and parameter set, which provides a quantitative measure of the degree of nonrelativistic localization.

Overall, our results reveal a unified picture of quarkonium LCDAs, characterized by longitudinal and transverse universality, nonrelativistic compactness, and an emergent twist-independence in the heavy-quark limit. These findings provide new insights into the nonperturbative structure of heavy quarkonium and offer a useful framework for future studies of hard exclusive processes involving heavy quark systems.

\section*{Acknowledgements}

Shuai Xu thanks the ICTP (Trieste) for its hospitality during the Summer School on Particle Physics. Qin Chang is supported by the National Natural Science Foundation of China (Grant No. 12275067), Science and Technology R$\&$D Program Joint Fund Project of Henan Province  (Grant No.225200810030), Science and Technology Innovation Leading Talent Support Program of Henan Province  (Grant No.254200510039), and National Key R$\&$D Program of China (Grant No.2023YFA1606000). Li-Li Chen is supported by the National Natural Science Foundation of China (Grant No.
12105078). Xiao-Nan Li is supported by the Anhui Provincial Department of Education Scientific Research Project (Grant No.2025AHGXZK40010) and the Tongling University Talent Program (Grant No.R23100). We would like to thank Xing-Gang Wu for valuable inspiration.


\begin{thebibliography}{99}

\bibitem{Quigg:1979vr}
C.~Quigg and J.~L.~Rosner,
Phys. Rept. \textbf{56}, 167-235 (1979).

\bibitem{Bodwin:1994jh}
G.~T.~Bodwin, E.~Braaten and G.~P.~Lepage,
Phys. Rev. D \textbf{51}, 1125-1171 (1995)

\bibitem{Bodwin:2006dn}
G.~T.~Bodwin, D.~Kang and J.~Lee,
Phys. Rev. D \textbf{74}, 014014 (2006)

\bibitem{Chernyak:1983ej}
V.~L.~Chernyak and A.~R.~Zhitnitsky,
Phys. Rept. \textbf{112}, 173 (1984)

\bibitem{Colangelo:2000dp}
P.~Colangelo and A.~Khodjamirian,
[arXiv:hep-ph/0010175 [hep-ph]].

\bibitem{Bakulev:2005cp}
A.~P.~Bakulev, S.~V.~Mikhailov and N.~G.~Stefanis,
Phys. Rev. D \textbf{73}, 056002 (2006)

\bibitem{Ball:1998sk}
P.~Ball, V.~M.~Braun, Y.~Koike and K.~Tanaka,
Nucl. Phys. B \textbf{529}, 323-382 (1998)

\bibitem{Yang:2007zt}
K.~C.~Yang,
Nucl. Phys. B \textbf{776}, 187-257 (2007)

\bibitem{Zhang:2025qmg}
S.~Q.~Zhang and C.~F.~Qiao,
[arXiv:2512.24706 [hep-ph]].

\bibitem{Zeng:2025rfe}
L.~Zeng, X.~G.~Wu, D.~D.~Hu, Y.~J.~Zhang, H.~B.~Fu and T.~Zhong,
Phys. Rev. D \textbf{111}, no.11, 116014 (2025)

\bibitem{Wang:2025sic}
Z.~G.~Wang,
Front. Phys. (Beijing) \textbf{21}, no.1, 016300 (2026)
\bibitem{Han:2013zg}
H.~Y.~Han, X.~G.~Wu, H.~B.~Fu, Q.~L.~Zhang and T.~Zhong,
Eur. Phys. J. A \textbf{49}, 78 (2013)

\bibitem{Zhong:2021epq}
T.~Zhong, Z.~H.~Zhu, H.~B.~Fu, X.~G.~Wu and T.~Huang,
Phys. Rev. D \textbf{104}, no.1, 016021 (2021)

\bibitem{Khodjamirian:2004ga}
A.~Khodjamirian, T.~Mannel and M.~Melcher,
Phys. Rev. D \textbf{70}, 094002 (2004)

\bibitem{Khodjamirian:2020hob}
A.~Khodjamirian, R.~Mandal and T.~Mannel,
JHEP \textbf{10}, 043 (2020)

\bibitem{Braguta:2008qe}
V.~V.~Braguta, A.~K.~Likhoded and A.~V.~Luchinsky,
Phys. Rev. D \textbf{79}, 074004 (2009)

\bibitem{Braguta:2007tq}
V.~V.~Braguta,
Phys. Rev. D \textbf{77}, 034026 (2008)

\bibitem{Braguta:2006wr}
V.~V.~Braguta, A.~K.~Likhoded and A.~V.~Luchinsky,
Phys. Lett. B \textbf{646}, 80-90 (2007)

\bibitem{Braguta:2007fh}
V.~V.~Braguta,
Phys. Rev. D \textbf{75}, 094016 (2007)

\bibitem{CP-PACS:2001vqx}
A.~Ali Khan \textit{et al.} [CP-PACS],
Phys. Rev. D \textbf{65}, 054505 (2002)

\bibitem{Braun:2006dg}
V.~M.~Braun, M.~Gockeler, R.~Horsley, H.~Perlt, D.~Pleiter, P.~E.~L.~Rakow, G.~Schierholz, A.~Schiller, W.~Schroers and H.~Stuben, \textit{et al.}
Phys. Rev. D \textbf{74}, 074501 (2006)

\bibitem{LatticeParton:2024vck}
M.~H.~Chu \textit{et al.} [Lattice Parton],
Phys. Rev. D \textbf{111}, no.3, 034510 (2025)

\bibitem{LatticeParton:2024zko}
X.~Y.~Han \textit{et al.} [Lattice Parton],
Phys. Rev. D \textbf{111}, no.3, 034503 (2025)

\bibitem{LatticeParton:2022zqc}
J.~Hua \textit{et al.} [Lattice Parton],
Phys. Rev. Lett. \textbf{129}, no.13, 132001 (2022)

\bibitem{Ding:2024saz}
H.~T.~Ding, X.~Gao, S.~Mukherjee, P.~Petreczky, Q.~Shi, S.~Syritsyn and Y.~Zhao,
JHEP \textbf{02}, 056 (2025)



\bibitem{Blossier:2024wyx}
B.~Blossier, M.~Mangin-Brinet, J.~M.~Morgado Ch{\'a}vez and T.~San Jos{\'e},
JHEP \textbf{09}, 079 (2024)

\bibitem{Zhang:2020gaj}
R.~Zhang, C.~Honkala, H.~W.~Lin and J.~W.~Chen,
Phys. Rev. D \textbf{102}, no.9, 094519 (2020)

\bibitem{Cloet:2024vbv}
I.~Cloet, X.~Gao, S.~Mukherjee, S.~Syritsyn, N.~Karthik, P.~Petreczky, R.~Zhang and Y.~Zhao,
Phys. Rev. D \textbf{110}, no.11, 114502 (2024)

\bibitem{Baker:2024zcd}
E.~Baker, D.~Bollweg, P.~Boyle, I.~Clo{\"e}t, X.~Gao, S.~Mukherjee, P.~Petreczky, R.~Zhang and Y.~Zhao,
JHEP \textbf{07}, 211 (2024)
\bibitem{Lepage:1980fj}
G.~P.~Lepage and S.~J.~Brodsky,
Phys. Rev. D \textbf{22}, 2157 (1980)

\bibitem{Chai:2025xuz}
J.~Chai and S.~Cheng,
JHEP \textbf{06}, 229 (2025)

\bibitem{Cheng:2019ruz}
S.~Cheng,
Phys. Rev. D \textbf{100}, no.1, 013007 (2019)

\bibitem{Cheng:2020vwr}
S.~Cheng, A.~Khodjamirian and A.~V.~Rusov,
Phys. Rev. D \textbf{102}, no.7, 074022 (2020)
\bibitem{Maris:1997tm}
P.~Maris and C.~D.~Roberts,
Phys. Rev. C \textbf{56}, 3369-3383 (1997)

\bibitem{Roberts:1994dr}
C.~D.~Roberts and A.~G.~Williams,
Prog. Part. Nucl. Phys. \textbf{33}, 477-575 (1994)

\bibitem{Chang:2013pq}
L.~Chang, I.~C.~Cloet, J.~J.~Cobos-Martinez, C.~D.~Roberts, S.~M.~Schmidt and P.~C.~Tandy,
Phys. Rev. Lett. \textbf{110}, no.13, 132001 (2013)

\bibitem{Chang:2013epa}
L.~Chang, C.~D.~Roberts and S.~M.~Schmidt,
Phys. Lett. B \textbf{727}, 255-259 (2013)

\bibitem{Shi:2015esa}
C.~Shi, C.~Chen, L.~Chang, C.~D.~Roberts, S.~M.~Schmidt and H.~S.~Zong,
Phys. Rev. D \textbf{92}, 014035 (2015)

\bibitem{Roberts:2021nhw}
C.~D.~Roberts, D.~G.~Richards, T.~Horn and L.~Chang,
Prog. Part. Nucl. Phys. \textbf{120}, 103883 (2021)

\bibitem{Xu:2025hjf}
Y.~Z.~Xu,
Phys. Rev. D \textbf{111}, no.11, 114012 (2025)

\bibitem{Ding:2015rkn}
M.~Ding, F.~Gao, L.~Chang, Y.~X.~Liu and C.~D.~Roberts,
Phys. Lett. B \textbf{753}, 330-335 (2016)
\bibitem{Petrov:1998kg}
V.~Y.~Petrov, M.~V.~Polyakov, R.~Ruskov, C.~Weiss and K.~Goeke,
Phys. Rev. D \textbf{59}, 114018 (1999)

\bibitem{Nam:2006au}
S.~i.~Nam, H.~C.~Kim, A.~Hosaka and M.~M.~Musakhanov,
Phys. Rev. D \textbf{74}, 014019 (2006)

\bibitem{Son:2024uet}
H.~D.~Son and P.~T.~P.~Hutauruk,
Phys. Rev. D \textbf{111}, no.5, 5 (2025)

\bibitem{Broniowski:2007si}
W.~Broniowski, E.~Ruiz Arriola and K.~Golec-Biernat,
Phys. Rev. D \textbf{77}, 034023 (2008)

\bibitem{RuizArriola:2002bp}
E.~Ruiz Arriola and W.~Broniowski,
Phys. Rev. D \textbf{66}, 094016 (2002)

\bibitem{Praszalowicz:2001wy}
M.~Praszalowicz and A.~Rostworowski,
Phys. Rev. D \textbf{64}, 074003 (2001)

\bibitem{Noguera:2015iia}
S.~Noguera and S.~Scopetta,
JHEP \textbf{11}, 102 (2015)

\bibitem{Courtoy:2019cxq}
A.~Courtoy, S.~Noguera and S.~Scopetta,
JHEP \textbf{12}, 045 (2019)

\bibitem{Choi:2007yu}
H.~M.~Choi and C.~R.~Ji,
Phys. Rev. D \textbf{75}, 034019 (2007)

\bibitem{Hwang:2008qi}
C.~W.~Hwang,
Eur. Phys. J. C \textbf{62}, 499-509 (2009)

\bibitem{Ji:1992yf}
C.~R.~Ji, P.~L.~Chung and S.~R.~Cotanch,
Phys. Rev. D \textbf{45}, 4214-4220 (1992)

\bibitem{Brodsky:1997de}
S.~J.~Brodsky, H.~C.~Pauli and S.~S.~Pinsky,
Phys. Rept. \textbf{301}, 299-486 (1998)

\bibitem{Terentev:1976jk}
M.~V.~Terentev,
Sov. J. Nucl. Phys. \textbf{24}, 106 (1976)

\bibitem{Chen:2021ywv}
L.~Chen, Y.~W.~Ren, L.~T.~Wang and Q.~Chang,
Eur. Phys. J. C \textbf{82}, no.5, 451 (2022)

\bibitem{Chang:2020wvs}
Q.~Chang, X.~L.~Wang and L.~T.~Wang,
Chin. Phys. C \textbf{44}, no.8, 083105 (2020)




\bibitem{Jaus1} W. Jaus, Phys. Rev. D {\bf 41}, 3394 (1990); Phys. Rev. D {\bf 44}, 2851 (1991).

\bibitem{CCH1} H. Y. Cheng, C. Y. Cheung and C. W. Hwang, Phys. Rev. D {\bf 55}, 1559 (1997).

\bibitem{Jaus2} W. Jaus, Phys. Rev. D {\bf 60}, 054026 (1999).

\bibitem{CCH2} H. Y. Cheng, C. K. Chua and C. W. Hwang, Phys. Rev. D {\bf 69}, 074025 (2004).

\bibitem{Hwang} C. W. Hwang, Phys. Rev. D {\bf 64}, 034011 (2001).

\bibitem{Wei} C. W. Hwang and Z. T. Wei, J. Phys. G {\bf 34}, 687
 (2007).
\bibitem{Hwang:2009cu}
C.~W.~Hwang,
JHEP \textbf{10}, 074 (2009)

\bibitem{Chang:2019mmh}
Q.~Chang, X.~N.~Li and L.~T.~Wang,
Eur. Phys. J. C \textbf{79}, no.5, 422 (2019)

\bibitem{Choi:2017uos}
H.~M.~Choi and C.~R.~Ji,
Phys. Rev. D \textbf{95}, no.5, 056002 (2017)

\bibitem{Dhiman:2019ddr}
N.~Dhiman, H.~Dahiya, C.~R.~Ji and H.~M.~Choi,
Phys. Rev. D \textbf{100}, no.1, 014026 (2019)

\bibitem{Arifi:2025olq}
A.~J.~Arifi, H.~M.~Choi and C.~R.~Ji,
Phys. Rev. D \textbf{112}, no.3, 033009 (2025)

\bibitem{Choi:2013mda}
H.~M.~Choi and C.~R.~Ji,
Phys. Rev. D \textbf{89}, no.3, 033011 (2014)


\bibitem{Chang:2019obq}
Q.~Chang, L.~T.~Wang and X.~N.~Li,
JHEP \textbf{12}, 102 (2019)

\bibitem{Choi:2025bxk}
Y.~Choi, A.~J.~Arifi, H.~M.~Choi and C.~R.~Ji,
[arXiv:2512.21642 [hep-ph]].

\bibitem{Zhang:2025pde}
F.~W.~Zhang and Z.~X.~Zhao,
[arXiv:2508.13648 [hep-ph]].

\bibitem{Xu:2025aow}
S.~Xu, X.~N.~Li and X.~G.~Wu,
Chin. Phys. Lett. \textbf{42}, no.8, 080201 (2025)

\bibitem{Xu:2025ntz}
S.~Xu, X.~N.~Li and X.~G.~Wu,
[arXiv:2507.07523 [hep-ph]].

\bibitem{Li:2026wmb}
X.~N.~Li, S.~Xu and Q.~Chang,
[arXiv:2601.01025 [hep-ph]].

\bibitem{Lih:2025cmf}
C.~C.~Lih and C.~Q.~Geng,
Phys. Rev. D \textbf{112}, no.7, 076023 (2025)

\bibitem{Lih:2026xgi}
C.~C.~Lih and C.~Q.~Geng,
[arXiv:2601.01124 [hep-ph]].

\bibitem{Wang:2024mjw}
Y.~L.~Wang, Y.~K.~Hsiao, K.~L.~Wang and C.~C.~Lih,
Phys. Rev. D \textbf{111}, no.9, 096013 (2025)

\bibitem{Hsiao:2020gtc}
Y.~K.~Hsiao, L.~Yang, C.~C.~Lih and S.~Y.~Tsai,
Eur. Phys. J. C \textbf{80}, no.11, 1066 (2020)

\bibitem{Arifi:2022pal}
A.~J.~Arifi, H.~M.~Choi, C.~R.~ji and Y.~Oh,
Phys. Rev. D \textbf{106}, no.1, 014009 (2022)

\bibitem{Bakamjian:1953kh}
B.~Bakamjian and L.~H.~Thomas,
Phys. Rev. \textbf{92}, 1300-1310 (1953)

\bibitem{Chung:1988my}
P.~L.~Chung, W.~N.~Polyzou, F.~Coester and B.~D.~Keister,
Phys. Rev. C \textbf{37}, 2000-2015 (1988)

\bibitem{Polyzou:2010zz}
W.~N.~Polyzou,
Phys. Rev. C \textbf{82}, 064001 (2010)
\bibitem{ParticleDataGroup:2022pth}
R.~L.~Workman \textit{et al.} [Particle Data Group],
PTEP \textbf{2022}, 083C01 (2022)

\bibitem{Wang:2020zbr}
G.~L.~Wang, T.~F.~Feng and X.~G.~Wu,
Phys. Rev. D \textbf{101}, no.11, 116011 (2020)
\bibitem{Zhong:2014fma}
T.~Zhong, X.~G.~Wu and T.~Huang,
Eur. Phys. J. C \textbf{75}, no.2, 45 (2015)

\bibitem{Zhong:2016kuv}
T.~Zhong, X.~G.~Wu, T.~Huang and H.~B.~Fu,
Eur. Phys. J. C \textbf{76}, no.9, 509 (2016)
\bibitem{Li:2017mlw}
Y.~Li, P.~Maris and J.~P.~Vary,
Phys. Rev. D \textbf{96}, 016022 (2017)



\bibitem{Brodsky:2014yha}
S.~J.~Brodsky, G.~F.~de Teramond, H.~G.~Dosch and J.~Erlich,
Phys. Rept. \textbf{584}, 1-105 (2015)




\end{thebibliography}
\end{document}